# Manipulating Fano coupling in an opto-thermoelectric field

Linhan Lin[1,†,*], Sergey Lepeshov[2,†], Alex Krasnok[3,*], Yu Huang[4], Taizhi Jiang[5], Xiaolei Peng[6], Brian A. Korgel[5,6], Andrea Alù[7,8,9], and Yuebing Zheng[6,10,*]

[1]State Key Laboratory of Precision Measurement Technology and Instruments, Department of Precision Instrument, Tsinghua University, Beijing 100084, People's Republic of China

[2]Department of Electrical and Photonics Engineering, DTU Electro, Technical University of Denmark, Building 343, DK-2800 Kgs. Lyngby, Denmark

[3]Department of Electrical and Computer Engineering, Florida International University, Miami, Florida 33174, USA

[4]School of Physics and Electronics, Hunan University, Changsha 410082, China

[5]Mc Ketta Department of Chemical Engineering, The University of Texas at Austin, Austin, TX 78712, USA

[6]Materials Science & Engineering Program and Texas Materials Institute, The University of Texas at Austin, Austin, TX 78712, USA

[7]Department of Electrical and Computer Engineering, The University of Texas at Austin, Austin, TX 78712, USA

[8]Photonics Initiative, Advanced Science Research Center, City University of New York, New York, NY 10031, USA

[9]Physics Program, Graduate Center, City University of New York, NY 10016, USA

[10]Walker Department of Mechanical Engineering, The University of Texas at Austin, Austin, TX 78712, USA

*E-mail: linlh2019@tsinghua.edu.cn; akrasnok@fiu.edu; zheng@austin.utexas.edu

†These authors contributed equally to this work.

**Abstract**

Fano resonances in photonics arise from the coupling and interference between two resonant modes in structures with broken symmetry. They feature an uneven and narrow and tunable lineshape, and are ideally suited for optical spectroscopy. Many Fano resonance structures have been suggested in nanophotonics over the last ten years, but reconfigurability and tailored design remain challenging. Herein, we propose an all-optical "pick-and-place" approach aimed at assemble Fano metamolecules of various geometries and compositions in a reconfigurable manner. We study their coupling behavior by in-situ dark-field scattering spectroscopy. Driven by a light-directed opto-thermoelectric field, silicon nanoparticles with high quality-factor Mie resonances (discrete states) and low-loss $BaTiO_3$ nanoparticles (continuum states) are assembled into all-dielectric heterodimers, where distinct Fano resonances are observed. The Fano parameter can be adjusted by changing the resonant frequency of the discrete states or the light polarization. We also show tunable coupling strength and multiple Fano resonances by altering the number of continuum states and discrete states in dielectric heterooligomers. Our work offers a general design rule for Fano resonance and an all-optical platform for controlling Fano coupling on demand.

Fano resonances, which were first explained by Ugo Fano as the interference between a discrete state and a continuum state in a quantum mechanical system, feature an asymmetric line shape in the absorption spectra of noble gases[1]. Fano resonances have been observed in a variety of quantum systems, such as quantum dots and nanowires[2]. This unusual interference phenomenon is not just found in quantum systems, it is also common in optics and photonics[3-6]. In nanophotonics, Fano resonances are associated with resonant optical phenomena, such as surface plasmons and Mie resonance, featuring a sharp transition at an extremely narrow frequency window in the optical spectroscopy (such as transmission, scattering, or absorption) [7-21]. Narrow Fano resonances have been harnessed to design different high-performance optical devices in optical sensing[22,23], nonlinearity[24-26], optical chirality[27], optical display[28], and more.

Fano resonances have been observed in various photonic nanostructures, including individual asymmetric nanostructures[9,29], nanoparticle assemblies[7,11,12,14,18,30], metamaterials[25-27,31,32], and photonic crystals[33-35]. In optics, the key to achieving Fano resonances is to combine a spectrally narrow mode with a much broader one. In homogenous clusters of nanoparticles, since each individual particle has a similar optical response, the generation of discrete states and continuum states relies on the mode hybridization, which complicates the structural design. A typical example is plasmonic nanoclusters consisting of packed metallic nanoscale components, i.e., a central particle surrounded by six equivalent particles[36]. The basic design rule is to obtain an equivalent dipole moment of the surrounding ring structure with the central particle, leading to a subradiant antibonding state coupled with a superradiant bonding state. The spectral feature of Fano resonance in these plasmonic clusters can be carefully designed by varying the composition, particle size, interparticle gap, and geometric arrangement[36-38]. However, real-time control over Fano resonances and the spectral lineshapes is still challenging.

From another perspective, the two modes in Fano interference can be created from individual components in the nanoclusters. Specifically, optical nanocavities made of different materials can be explored to control Fano interference. It has been demonstrated that electrostatically assembled Au-Ag nanodimers exhibit Fano resonance in the absorption spectra, indicating the possibility of making photonic Fano nanostructures through bottom-up assembly[16]. The nanoparticles were labeled with different surface charges, allowing selective interaction during the assembly. However, the variety and spread of the nanoparticles complicates the precise control of geometry and optical response of the nanocluster. It is still an appealing perspective to selectively pick up nanoparticles with different optical properties and assemble them into photonic nanoclusters at will to investigate the coupling behavior.

To address this challenge, we propose an all-optical approach to "pick-and-place" different nanoparticles into hetero-metamolecules, with their coupling behavior investigated using *in-situ* spectroscopy. Technically, a focused laser beam is used to heat a thin Au film, creating a thermal hot spot due to optical heating. Cetyltrimethylammonium chloride (CTAC) surfactant was added into the nanoparticle suspension, providing the macro cations (CTAC micelles) and counter $Cl^-$ ions to create a light-directed thermoelectric field, which captures the surfactant-adsorbed nanoparticles at the laser spot[39]. Additionally, the CTAC micelles also act as the ionic depletants and provide the depletion attraction force for the interparticle bonding[40,41]. By steering the laser beam, nanoparticles can be dynamically moved and assembled into various nanopatterns[42,43]. As a first demonstration, we organized 500 nm SiNPs into diverse oligomers, with the particle number ranging from one to ten (Fig. 1a). The opto-thermoelectric field allows to manipulate nanoparticles of different sizes (see Supplementary Fig. S1 for the assembly of 300 nm SiNPs) and the reconfigurable geometric control of the assembled nanoclusters (see Supplementary Fig. S2).

To achieve Fano coupling, we choose BaTiO$_3$ nanoparticles (NPs) which exhibit broadband scattering (continuum state), and amorphous silicon particles (SiNPs) of different sizes which show multiple high quality-factor Mie modes (discrete state), as the building blocks. We use in-situ dark-field scattering spectroscopy to identify different nanoparticles and analyze their coupling behavior before and after assembly. More interestingly, the reconfigurable assembly of these hetero-metamolecules allows real-time tuning of the cluster structure to manipulate the coupling behavior in the opto-thermoelectric trap. This versatile optical technique allows to build diverse Fano metamolecules to investigate the underlying coupling physics, which will in turn guide the design of Fano nanostructures with tailorable optical properties.

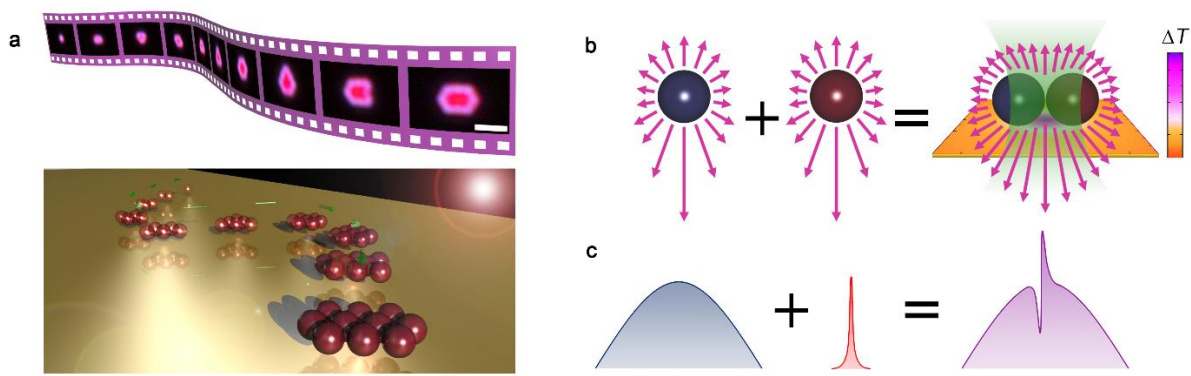

**Figure 1. Design concept and working principle. a,** Scheme (bottom) and dark-field optical images (top) showing the successive assembly of 500 nm SiNPs into diverse oligomers. Scale bar: 2 μm. **b and c,** Design concept of the Fano heterodimer. Two individual particles, one exhibiting broad scattering spectrum while another exhibiting narrow scattering spectrum, are assembled into a heterodimer using a light-controlled temperature field. The arrows represent the scattering signal of the particles.

**Manipulating Fano heterodimers**

Fig. 2a shows the scheme and dark-field optical images of a single 300 nm $BaTiO_3$ particle. The single $BaTiO_3$ particle appears white under dark field microscope, matching the broadband scattering spectrum from visible to near-infrared (Fig. 2b). The decomposition of scattering spectra reveals that the broad optical response arises from the magnetic dipole ($M_1$) and electric dipole ($E_1$) modes (Supplementary Fig. 3). Unlike $BaTiO_3$, Si has a high refractive index (above 3.5) in the visible and near-infrared regions (300 nm to 800 nm), making sub-micron Si particles excellent Mie resonators. Besides, the spherical shape and smooth surface of the SiNPs improve their quality factor. Fig. 2b shows the scattering spectra of individual SiNPs with diameters ranging from 420 nm to 476 nm. High-quality-factor scattering peaks were observed in each particle, with all the modes red-shift successively when the particle size increases. Scattering intensity below 600 nm is low due to the high optical absorption of amorphous Si, matching the red color in the dark-field optical image.

When SiNPs and $BaTiO_3$ particles formed heterodimers, multiple asymmetric dips appeared in the scattering spectra, indicating Fano coupling. We used polarization-dependent dark-field spectroscopy to study the coupling behavior in-situ. For a dimer consisting of a 300 nm $BaTiO_3$ particle and a 420 nm SiNP, two Fano dips appeared at 680 nm and 782 nm when the light polarization was perpendicular to the dimer axis. We studied the spectral evolution by replacing the 420 nm SiNP with SiNPs of different diameters in the heterodimer, as shown in Fig. 2c. We can see a clear redshift of the Fano dips when the size of SiNPs increases ($FR_1^\perp$ and $FR_2^\perp$ indicated by the red dashed curve and green dashed curve, respectively).

Three spectral features were noted. First, the Fano resonance frequencies were almost the same as the scattering peaks in individual SiNPs. Second, Fano coupling occurred between 650 nm and 835 nm (the colored region). New Fano resonances are observed when the high-quality-factor scattering peaks of individual SiNPs enter this regime ($FR_3^\perp$ indicated by the blue dashed curve). The multipole decomposition of the scattering spectrum of individual $BaTiO_3$ particle (Supplementary Fig. 3) suggests that the continuum state stems from $M_1$ mode instead of $E_1$ mode. Third, the Fano parameter is tunable by controlling the SiNP size in the heterodimers.

The orientation of electromagnetic (EM) field in the Mie resonators is sensitive to the polarization of incident light, which significantly modifies the interparticle interaction. When the light polarization is parallel to the dimer axis, $FR_3^{||}$ exhibits similar spectral evolution as $FR_3^\perp$. However, distinct coupling behavior is observed at other wavelengths. For instance, $FR_1^{||}$ shows similar redshift with the increase of SiNP size, while the Fano parameter is totally different from that of $FR_1^\perp$. Moreover, $FR_2^{||}$ is not observed for parallel polarization, where the scattering peaks similar to the ones of individual SiNPs are observed.

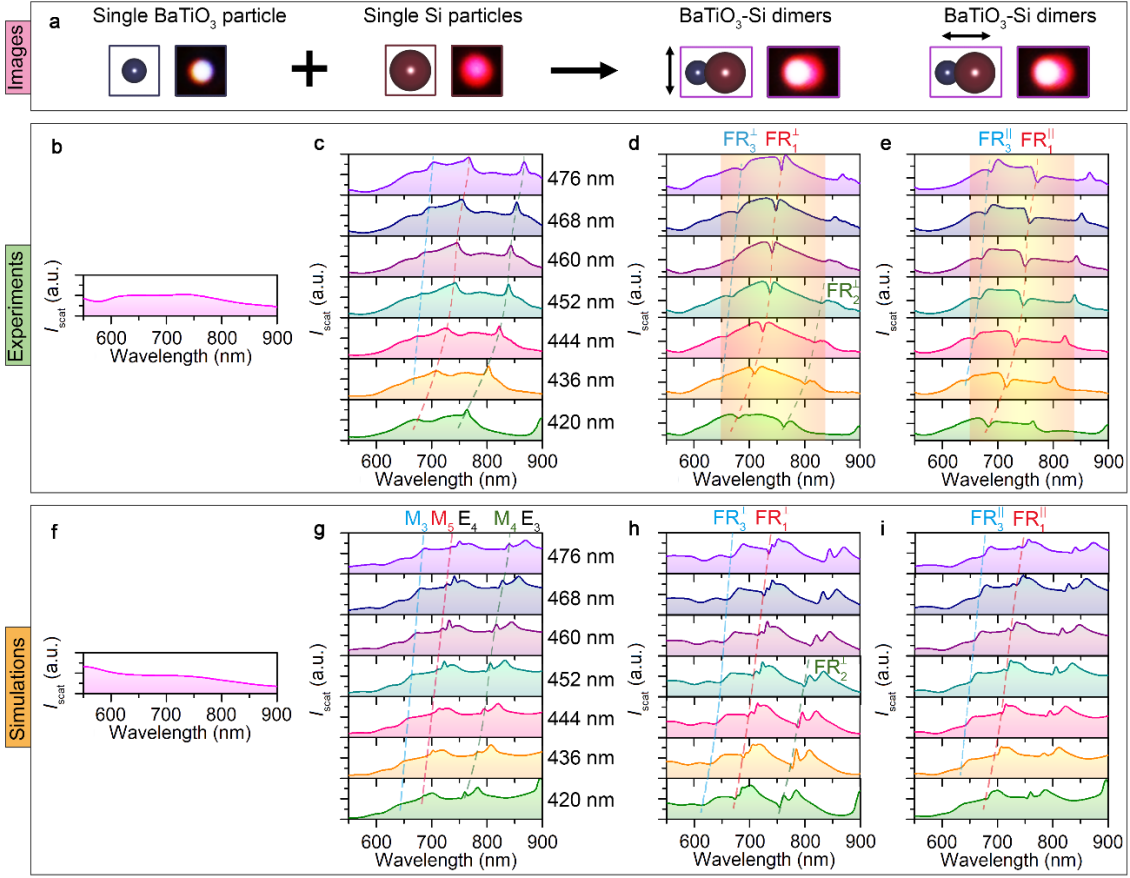

**Figure 2. Fano coupling in BaTiO$_3$-SiNP Heterodimers. a,** Scheme and dark-field optical images of a single BaTiO$_3$ NP, a single SiNP, and a BaTiO3-Si heterodimer (under perpendicular and parallel excitation). **b-e,** Experimental scattering spectrum of a single BaTiO$_3$ NP, single SiNPs with diameters from 420 nm to 476 nm, BaTiO3-Si heterodimer at perpendicular polarization, BaTiO3-Si heterodimer at parallel polarization, respectively. **f-i,** Simulated scattering spectrum of a single BaTiO$_3$ NP, single SiNPs with diameters from 420 nm to 476 nm, BaTiO3-Si heterodimers at perpendicular polarization, BaTiO$_3$-Si heterodimers at parallel polarization, respectively. Dashed curves show the evolution of scattering modes in SiNPs and their coupling with the magnetic dipole (M$_1$) mode in BaTiO$_3$ particles. The electrical (E$_l$) and magnetic (M$_l$) spherical Mie modes of l orders are marked in **g**: l = 1 (dipole), l = 2 (quadrupole), l = 3 (octupole), l = 4 (hexadecapole), etc. FR$_n^\perp$ and FR$_n^\parallel$ (as well as the dashed curves with the same color) means

different sets of Fano resonances at perpendicular polarization and parallel polarization, respectively. The colored rectangle gives the wavelength range of the magnetic dipole mode ($M_1$) of individual $BaTiO_3$ particle.

To understand the coupling behavior of Fano resonances observed in the heterodimers, we simulated the scattering spectra of individual $BaTiO_3$ particle, individual SiNPs, and heterodimers at different light polarizations, respectively, as summarized in Figs. 2f-i. The simulated results show similar spectral features to the experimental ones, with additional scattering peaks observed. These additional peaks arise from high-order Mie modes in the SiNPs, while they are difficult to detect in experiments due to the imperfect spherical shapes. From the simulation of heterodimers, $FR_2^\perp$ (green dashed curve in Fig. 2g) arises from the coupling between $M_4$ mode in SiNPs and $M_1$ mode in $BaTiO_3$ particle at perpendicular polarization. The Fano lineshape disappears when the SiNP size exceeds 460 nm as there is no spectral overlap between these two modes. When the light polarization is parallel to the dimer axis, the Fano coupling did not occur, which is consistent with the experimental observation. At the short wavelength, both $FR_3^\perp$ and $FR_3^\parallel$ (see the blue dashed curves in Figs. 2g and 2h) arise from the coupling between $M_3$ modes in SiNPs and $M_1$ modes in $BaTiO_3$ particle, which explain the similar spectral evolution at both polarizations. For $FR_1$, the polarization-dependent Fano coupling behavior arises from different mode couplings, i.e., $FR_1^\perp$ is caused by the coupling between $M_5$ mode of SiNPs and $M_1$ mode of $BaTiO_3$ particle, and $FR_1^\parallel$ is generated by the coupling between $E_4$ mode of SiNPs and $M_1$ mode of $BaTiO_3$ particle.

To further understand the polarization-dependent Fano coupling behavior, we measured the dark-field scattering spectra of the 300 nm $BaTiO_3$-476 nm SiNP heterodimer at different incident light polarizations (see the experimental setup in Fig. 3a). As shown in Fig. 3b, the dip wavelength of $FR_3$ is almost independent on the light polarization, while the Fano parameter changes successively as a function of the polarization angle, indicating similar mode interference at different polarizations. In contrast, $FR_1$ exhibits distinct polarization-dependent behavior. A Fano dip at 754 nm is seen at 90 degrees but becomes inconspicuous and disappears at 0 degrees. Another dip at 765 nm appears and becomes arresting when the polarization degree decreases. At 60 degrees, both Fano dips at 754 nm and 765 nm coexist, confirming two different mode interferences for $FR_1$: $M_5$ mode of SiNPs coupling with $M_1$ mode of $BaTiO_3$ at 90 degrees, and $E_4$ mode of SiNPs coupling with $M_1$ mode of $BaTiO_3$ at 0 degrees. The wavelength offset between the two Fano dips of $FR_1$ depends on the SiNP size (see top panel of Fig. 3c), explained by the offset between $E_4$ and $M_5$ modes in single SiNP spectra (see bottom panel of Fig. 3c).

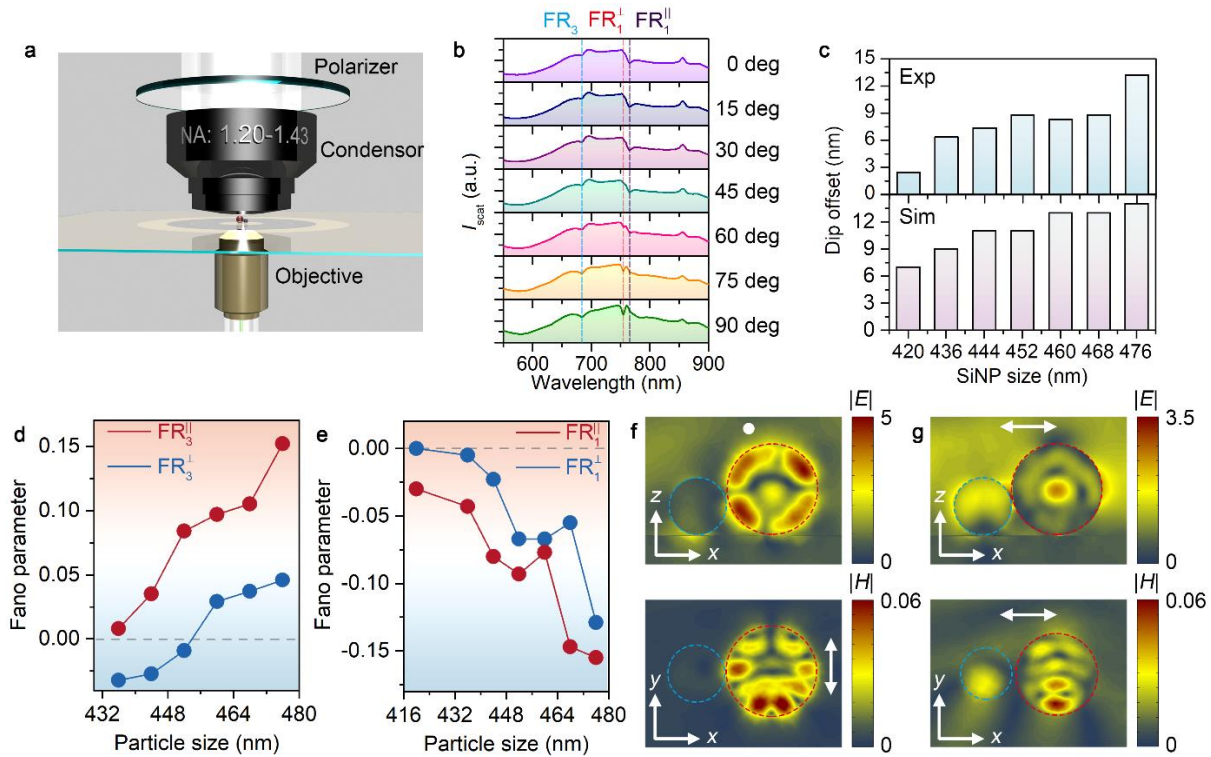

**Figure 3. Polarization effects and tunable Fano parameter in 300 nm $BaTiO_3$-500 nm SiNP heterodimers. a,** Scheme showing polarization-dependent dark-field scattering spectroscopy. **b,** Polarization-dependent scattering spectra of 300 nm $BaTiO_3$-500 nm SiNP heterodimers. The blue dashed line represents the evolution of dip location for $FR_3$ when the polarization angle changes

from 0 to 90 degree. The blue, red, and violet dashed lines show the resonance frequencies of $FR_3$, $FR_1^{\perp}$, and $FR_1^{\parallel}$, respectively. **c,** Wavelength offset of Fano resonances $FR_1$ between perpendicular and parallel polarizations at different SiNP sizes (experiment, top); Wavelength offset between $E_4$ mode and $M_5$ mode in single SiNP (simulation, bottom). **d,** Fano parameter as a function of SiNP size at different polarizations for $FR_3$. **e,** Fano parameter as a function of SiNP size at different polarizations for $FR_1$. **f,** Side view of electric field distribution (top) and top view of magnetic field distribution (bottom) for $FR_1^{\perp}$. **g,** Side view of electric field distribution (top) and top view of magnetic field distribution (bottom) for $FR_1^{\parallel}$. White arrows indicate light polarization.

To further verify the Fano resonance in the $BaTiO_3$-Si heterodimers, we fit the experimental scattering spectra using an analytical function $\sigma_t(\omega) = \sigma_d(\omega)\sigma_c(\omega)$, where the terms of discrete state ($\sigma_d(\omega)$) and continuum state ($\sigma_c(\omega)$) can be described as

$$\sigma_d(\omega) = \frac{\left(\frac{\omega^2-\omega_d^2}{2\gamma_d\omega_d}+q\right)^2+b}{\left(\frac{\omega^2-\omega_d^2}{2\gamma_d\omega_d}\right)^2+1} \tag{1}$$

$$\sigma_c(\omega) = \frac{a^2}{\left(\frac{\omega^2-\omega_c^2}{2\gamma_c\omega_c}\right)^2+1} \tag{2}$$

In these equations, $\omega_d$ and $\omega_c$ are the central frequencies, and $\gamma_d$ and $\gamma_c$ are the spectral linewidths of the discrete and continuum states, respectively. $b$ is the damping parameter from intrinsic losses, $a$ is the maximum resonance amplitude, and $q$ is the asymmetric Fano parameter. The asymmetric Fano parameter as a function of the SiNP size for $FR_3^{\perp}$ and $FR_3^{\parallel}$ is summarized in Figs. 3e. The $q$ value of $FR_3^{\parallel}$ is always positive and it increases monotonously with the SiNP size. In contrast, the $q$ value of $FR_3^{\perp}$ is switched from negative to positive when the SiNP size exceeds 460 nm. The $q$ value of both $FR_1^{\parallel}$ and $FR_1^{\perp}$ decreases with the SiNP size, while a maximal value is observed when the SiNP size is 460 nm or 468 nm. We examine the EM field distribution of 300 nm BaTiO$_3$-500 nm SiNP heterodimers for FR$_1$. At perpendicular polarization, the SiNP shows an EM field of M$_5$ mode, with symmetry interrupted by interparticle interference ( see Supplementary Fig. 4). The BaTiO$_3$ field is hardly visible due to height differences. At parallel polarization, the electric field on the SiNP surface is observed, which couples with the BaTiO$_3$ particle.

The Mie resonance of SiNPs is tunable by controlling their size. Electric and magnetic modes can be tuned to match the magnetic dipole mode of BaTiO$_3$ particles for Fano coupling. We assembled heterodimers with SiNPs of 300 nm and 700 nm in diameters, as shown in Supplementary Fig. 5. In 300 nm BaTiO$_3$-300 nm SiNP heterodimers, low-order electric and magnetic modes appear in the visible and near-infrared regime. We observed the coupling between E$_2$ mode of the 300 nm SiNPs and M$_1$ mode of the BaTiO$_3$ particle, leading to Fano resonance at both polarizations. However, we cannot see the Fano interference between the M$_2$ mode of the SiNPs and M$_1$ mode of the BaTiO$_3$ particle, while a suppression of scattering intensity is observed (especially for the perpendicular polarization). The Fano resonance design can also be applied to heterodimers with larger SiNPs, such as 700 nm. As shown in Supplementary Fig. 5, the scattering peaks between 650 nm and 835 nm in single SiNP spectra evolve into Fano dips once they are assembled with BaTiO$_3$ particle. However, the scattering peaks of single 700 nm SiNPs in this range stem from high-order modes, making simulation challenging to explain the coupling behavior.

**Manipulating Fano heterooligomers**

Understanding coupling behavior in BaTiO$_3$-SiNP Fano heterodimers guides the design and manipulation of Fano resonance in more complex structures. BaTiO$_3$ particles act as continuum states and SiNPs as discrete states, allowing us to tailor Fano resonances by varying their numbers. We expect that adding $BaTiO_3$ particles increases the density of continuum states and improves coupling strength. As a proof-of-concept, we assembled heterooligomers with a 500 nm SiNP and $BaTiO_3$ particles of different number (from one to five). The coupling between the SiNP and each $BaTiO_3$ particle is similar to each other (Supplementary Fig. 6). The evolution of scattering spectra verifies our hypothesis, i.e., the Fano dips for both $FR_1$ and $FR_3$ become deeper when the number of $BaTiO_3$ particles increase, suggesting improvement of the coupling strength. For the heterooligomer with five $BaTiO_3$ particles, coupling strength is slightly reduced due to the imperfect shape of $BaTiO_3$ particles and the structural instability of the metamolecules. Moreover, the addition of $BaTiO_3$ particles induces near-field coupling among the $BaTiO_3$ particles and expands the working range of continuum states, providing possibilities to couple with the discrete states beyond 835 nm. In the heterodimer, a scattering peak at 864 nm from the SiNP's $M_4$ mode cannot couple with the $BaTiO_3$ particle's $M_1$ mode due to lack of spectral overlap. In the heterooligomer with three $BaTiO_3$ particles, the peak becomes a Fano dip labeled as $FR_2$. Increasing the number of $BaTiO_3$ particles further improves coupling strength. Positive and decreasing $q$ values are also obtained with the increased number of $BaTiO_3$ particles. The observed coupling behavior well matches the simulation (Fig. 4b). The EM field distribution of $FR_1$ from the heterooligomers (Figs. 4c and d) further verifies the physical origin of the discrete state, i.e., $M_5$ mode in the SiNP.

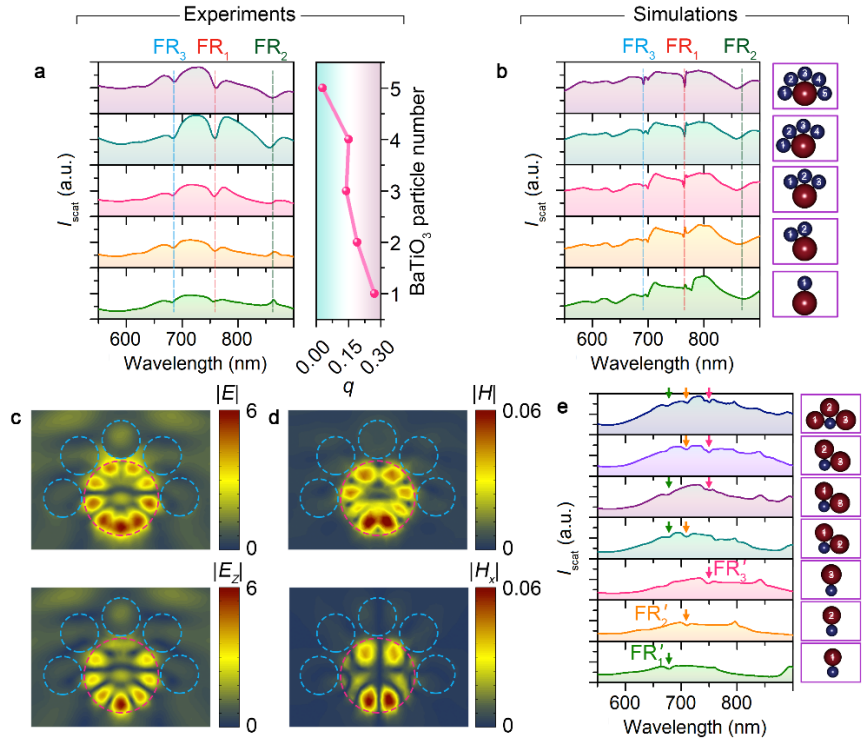

**Figure 4. Manipulation of Fano resonances in BaTiO$_3$-SiNP heterooligomers. a,** Experimental scattering spectra of heterooligomers with a 500 nm SiNP and varying numbers of BaTiO$_3$ particles (bottom to top: one to five, see left panel) and the asymmetric Fano parameter $q$ as a function of BaTiO$_3$ particles (right panel). **b,** Simulated scattering spectra of heterooligomers with a 500 nm SiNP and varying numbers of BaTiO$_3$ particles. Geometries are shown in the right panels. **c,** Top view of the electric field distribution (top: $|E|$, bottom: $|E_z|$) of FR$_1$ for BaTiO$_3$-SiNP heterooligomers consisting of 5 BaTiO$_3$ particles. **d,** Top view of the magnetic field distribution (top: $|H|$, bottom: $|H_x|$) of FR$_1$ for BaTiO$_3$-SiNP heterooligomers consisting of five BaTiO$_3$ particles. **e,** Experimental scattering spectra of heterooligomers consisting of a 300 nm BaTiO$_3$ particle and different number of SiNPs (one to three). The arrows with different color indicate the Fano resonance induced by different SiNPs, i.e., FR$_1'$ by SiNP$_1$, FR$_2'$ by SiNP$_2$, and FR$_3'$ by SiNP$_3$. Geometries are shown in the right panels.

Besides the engineering of continuum states in the heterooligomers, we anticipate that selection and addition of SiNPs into the heterooligomers provides another degree of freedom to tailor the Fano resonance. For verification, we picked up a $BaTiO_3$ particle and three SiNPs with different sizes (labeled as $SiNP_1$, $SiNP_2$, and $SiNP_3$), with the single-particle scattering spectra summarized in Supplementary Fig. 7. First, we assembled each SiNP with the $BaTiO_3$ particle to form three different heterodimers and recorded their scattering spectra, respectively. The Fano resonance arising from the coupling between $M_5$ mode of the SiNP and $M_1$ mode of the $BaTiO_3$ particle in these heterodimers are written as $FR'_1$, $FR'_2$, and $FR'_3$, respectively. Then, we assembled the $BaTiO_3$ particle with two of the SiNPs to form three types of heterotrimers. Interestingly, we can see clearly an add-up effect. That means, when $SiNP_i$ and $SiNP_j$ ($i, j$=1, 2, or 3) are assembled with the $BaTiO_3$ particle to form a trimer, the Fano resonance in both heterodimers ($FR'_i$ and $FR'_j$) can be observed in the trimer. Moreover, we assembled the four particles together to form a heterotetramer and observed that $FR'_1$, $FR'_2$, and $FR'_3$ co-exist in the scattering spectrum. This add-up effect provides the opportunities to design versatile and multiple Fano resonance.

**Conclusions**

Using a light-directed thermoelectric field, we demonstrated the dynamic manipulation of Fano coupling in an optical trap. Dielectric nanoresonators with distinct optical properties are trapped and assembled into heterodimers or heterooligomers to study Fano interference. Unlike the Fano theory based on mode hybridization, we found that individual dielectric particles support continuum or discrete states, allowing precise design and control of Fano interference through selective assembly. Using SiNPs (discrete states) and $BaTiO_3$ particles (continuum states) as examples, we successfully tuned the Fano coupling by controlling particle size, composition, and light polarization. Moreover, we observe an interesting add-up effect in the BaTiO$_3$-SiNP heterooligomers, which allow to design multiple Fano resonances in a versatile manner. The on-demand manipulation of Fano resonance in this work provides opportunities to gain insight of the Fano coupling theory, which will find a lot of applications in active photonics and bio-sensing.

**Acknowledgement**

L.L. acknowledges support from the National Key Research and Development Program of China (grant 2020YFA0715000), the National Natural Science Foundation of China (grant 62075111), and the Tsinghua University Initiative Scientific Research Program. Y. Z. acknowledge the financial support of the National Institute of General Medical Sciences of the National Institutes of Health (R01GM146962).